\documentstyle[epsfig]{mn}

\input{psfig.sty}

\ifnfsstwo

\fi

\ifnfssone
  \newmathalphabet{\mathit}
    \addtoversion{normal}{\mathit}{cmr}{m}{it}
    \addtoversion{bold}{\mathit}{cmr}{bx}{it}

\fi

\ifoldfss

\fi

\loadboldmathitalic
\loadboldgreek

\title[Chemical Evolution of dSphs and BCGs]
       {Chemical Evolution of Dwarf Spheroidal and Blue Compact Galaxies}
\author[G. A. Lanfranchi and F. Matteucci]
 {Gustavo A. Lanfranchi $^{1, 2}$ and Francesca Matteucci$^2$\\
$^1$IAG-USP,
 R. do Mat\~ao 1226, Cidade Universit\'aria, 05508-900 S\~ao Paulo, 
SP, Brazil\\
 $^2$ Dipartimento di Astronomia-Universit\'a di Trieste,
  Via G. B. Tiepolo 11, 34131 Trieste, Italy}

\pubyear{2002}

\begin{document}
\maketitle

\begin{abstract}
We studied the star formation and chemical evolution in a sample 
of 8 Dwarf Spheroidal (dSph) galaxies of the Local Group and in
Blue Compact Galaxies (BCGs) by means of comparison between
the predictions of chemical evolution models and several observed
abundance ratios. Detailed models with up to date nucleosynthesis 
taking into account the role played by supernovae of 
different types (II, Ia) were developed for both types of galaxies
allowing us to follow the evolution of several chemical elements 
(H, D, He, C, N, O, Mg, Si, S, Ca, and Fe). The models are specified
by the prescriptions of the star formation and galactic wind
efficiencies chosen to reproduce the main features of these 
galaxies. The BCGs are characterized by a star
formation proceeding in several short bursts separated by long 
quiescent periods and by a low wind efficiency, whereas one or two 
long bursts and a  very efficient wind well describe the dSph 
galaxies. We also investigated a possible connection in the
evolution of dSph and BCGs and compared the predictions
of the models to the abundance ratios observed in Damped Lyman 
$\alpha$ Systems (DLAs). The main conclusions are: i) the observed 
distribution of [$\alpha$/Fe] vs. [Fe/H] in dSph galaxies is 
mainly a result of the star formation rate coupled with 
the wind efficiency; ii) a low star formation efficiency
($\nu=0.01 - 1\,Gyr^{-1}$) and a high wind efficiency 
($w_i \sim 5-15$) are required to reproduce 
the observational data for dSph galaxies; iii) the low gas content
of these galaxies is the result of the combined action of gas 
consumption by star formation and gas removal by galactic winds; 
iv) the BCGs abundance ratios are reproduced by models with 2 to 7
bursts of star formation and an efficiency in the range 
$\nu=0.1 - 0.9\, Gyr^{-1}$; v) the low values of N/O observed 
in BCGs are the natural result of
a bursting star formation; vi) a connection between dSph and 
BCGs in an unified evolutionary scenario is unlikely;
vii) the models for both the dSph galaxies and BCGs reproduce
the abundance ratios observed in DLAs, but imply different 
formation scenarios for these objects; viii) a suitable amount of primary N 
produced in massive stars can be perhaps an explanation for the low 
plateau in the [N/$\alpha$] distribution observed in DLAs, if real.
\end{abstract}

\begin{keywords}
galaxies: abundances -- galaxies: Local Group --
galaxies: evolution -- galaxies: high-redshift -- quasars: general
\end{keywords}

\section{Introduction}

The Dwarf Spheroidal galaxies are important objects 
to understand the formation and evolution of galaxies. 
They could be the building blocks for the assembly of the 
larger galaxies in the scenario of hierarchical structure 
formation and contribute to improve our knowledge of the 
nucleosynthesis of the elements. Understanding their 
star formation (SF) and chemical enrichment
histories can help in clarifying these issues. If the Local
Group dSph galaxies are indeed the small structures from
which the halo of our Galaxy was formed, one should see a 
similarity between the patterns of the abundance ratios
observed in the halo stars and in the dSph galaxies.
Besides that, the determination of the abundances and 
abundance ratios of several 
elements in these galaxies offers an unique opportunity to 
improve our knowledge on the formation of the elements.

Recently, the dSph galaxies have been the subject of 
a great deal of studies concerning accurate abundance determination
with high dispersion spectroscopy allowing a detailed
study of their star formation and chemical enrichment histories
(Shetrone, Bolte \& Stetson 1998; Smecker-Hane \&
McWilliam 1999;  Bonifacio et al. 2000; Shetrone, Cot\'e 
\& Sargent 2001; Shetrone et al. 2003; Tolstoy et al. 2003).
Shetrone et al. (2003), using UVES spectra of 15 individual red 
giants in 4 dSph galaxies of the Local Group, measured the 
abundances of a series of elements including $\alpha$ and iron-peak 
ones. They showed that the dSph giants exhibit $\alpha$/Fe ratios
lower than those observed in the metal-poor Galactic halo in the 
same abundance range, as already noticed by Shetrone, Cot\'e \& 
Sargent (2001). They suggested that the low abundance ratios at low 
metallicities could be due to a slow star formation rate (SFR),
which was also suggested by Tolstoy et al. (2003) for the 
same galaxies. Tolstoy et al. (2003) also noticed that the dSph 
galaxies of the Local Group exhibit 
similar abundance patterns suggesting a fairly uniform chemical 
evolution, despite their different 
star formation histories. 

The star formation histories of dSph galaxies of the Local
Group were inferred by means of color magnitude diagrams (CMDs). 
These galaxies show a wide range of different star formation 
histories with some systems showing signs of extended or bursting
periods of star formation while others are consistent with 
just one burst in the early epochs of the evolution of the galaxy
(Smecker-Hane 1997; Smecker-Hane \& McWilliam 1999;
Hernandez, Gilmore \& Valls-Gabaud 2000; Dolphin 2002). 
These differences in the star formation histories should be 
visible also in the pattern of the observed abundance ratios, 
especially of those elements which are produced on different 
time-scales, such as the $\alpha$ and iron-peak elements, 
respectively. Thus, the comparison of key abundance ratios 
with predictions of chemical evolution models allows one to 
constrain the effects of star formation episodes and different 
star formation histories.

Models for dSph galaxies using star formation histories 
and the observed [Mg/Fe] as observational constraints were 
proposed recently in the literature (Carraro et al. 2001;
Carigi, Hernandez \& Gilmore 2002; Ikuta \& Arimoto 2002).
Ikuta \& Arimoto (2002) used a chemical evolution model
coupled with a simulation code for CMD in order to calculate
the star formation and chemical enrichment histories considering 
the effects of the chemical evolution on photometric properties. 
They concluded that the observed [Mg/Fe] ratios and the 
morphologies of the observed CMDs of local dSph 
galaxies can be reproduced by a star formation with long duration
($>$ 3.9 - 6.5 Gyr) and very low rates (10$^{-2}$ - 5 $\cdot 10^{-3}$
Gyr$^{-1}$). In a different approach, Carigi, Hernandez $\&$ 
Gilmore (2002) used star formation histories inferred by CMDs as 
a constraint to obtain information on the history of parameters of 
galactic evolution such as gas infall, outflows and global 
metallicities of local dSph galaxies. Carraro et al. (2001) using 
N-body/hydro-dynamical simulations including star formation, 
supernova (SN) feed-back and chemical evolution suggested that 
there is an ample possibility for the star formation history in 
these galaxies, ranging from a single short initial episode to 
longer ones, burst-like activities and very prolonged star 
formation in agreement with what is suggested by CMDs. In their
simulations the gas content is removed from the luminous part 
of the galaxy by galactic winds.

The galactic wind, in fact, might be a very important feature in 
the evolution of the dSph galaxies since these are very gas-poor 
systems. Indeed, in many of these galaxies no interstellar
medium (ISM) was detected and
upper limits on the ratio between HI and total mass are
around 0.004 (Mateo 1998). However, the mechanism responsible for 
the removal of the gas content of these galaxies is not known 
and some assumptions such as gas stripping by our Galaxy 
(Ikuta $\&$ Arimoto 2002) or winds triggered by supernova 
explosions (Vader 1986; Matteucci $\&$ Tornamb\'e 1987;
Arimoto $\&$ Yoshii 1987; Carraro et al. 2002; Silk 2002) are
required. 

Another issue concerning the dSph galaxies is the suggestion
of a connection between these gas-poor galaxies and late-type 
gas-rich galaxies (Blue Compact Galaxies - BCGs) in an unified 
evolutionary scenario, in the sense that one type (BCGs) could 
evolve into the other (dSph) by means of the consumption 
of the gas into stars by SF or removal of gas by ram 
pressure stripping (Lin $\&$ Faber 1983; Dekel $\&$ Silk 1986;
Davies $\&$ Phillipps 1988; van den Bergh 1994;
Papaderos et al. 1996a).

The BCGs (also called HII Galaxies) are small galaxies
with high central surface brightness, large content of 
neutral gas (up to $\sim$ 10 times the stellar content), 
low metallicities (from 0.5 Z$_{\odot}$ to 0.02 Z$_{\odot}$), 
and ongoing star formation. All these properties lead to the 
idea that the BCGs should be poorly evolved objects either 
young or which have undergone few episodes of star 
formation separated by long quiescent periods. The last hypothesis 
seems to be the most likely for the majority of the BCGs 
(see Kunth $\&$ \"Ostlin 2000 for a review).

A great deal of theoretical models for BCGs were proposed in 
order to explain the observed abundances (Matteucci $\&$ Chiosi 
1983; Matteucci $\&$ Tosi 1985; Pilyugin 1993; Carigi et al. 1994;
Marconi, Matteucci $\&$ Tosi 1994; Kunth, Matteucci $\&$ Marconi
1995). They all derived similar scenarios for the chemical evolution
in these systems: the IMF is in general the same everywhere with 
a slope similar to the Salpeter (1955) one, galactic winds with 
varying efficiencies are necessary to explain the observational
data, and the star formation is discontinuous and 
characterized by short bursts separated by quiescent periods.
The same scenario was adopted by Bradamante, Matteucci $\&$
D'Ercole (1998 - hereafter BMD)
who proposed a model with short and intense bursts of 
activity where the stellar energetics and the presence 
of dark matter were taken properly into account in order 
to explain the observed N/O, C/O and O/Fe ratios. 
Their model suggested that the number of bursts should be
between 1 and 10 with an efficiency of star formation varying from
0.1 to 7 Gyr $^{-1}$ and a Salpeter IMF. They also concluded 
that the BCGs should be dominated by dark 
matter, with a ratio between dark and luminous matter R = 1-50
in order to be gravitationally bound against the intense 
starbursts, and that galactic winds are differential, in the sense
that is likely that some chemical elements are preferentially lost 
from the system.

More recently, Recchi et al. (2001, 2002), analysed the
effects of starbursts on the dynamical and chemical evolution 
of BCGs, by taking into account, in detail, the chemical and 
energetical contributions from type II (SNe II) and type Ia 
(SNe Ia) supernovae. The dynamical and chemical evolution of the 
ISM was followed by means of a two-dimensional
hydrodynamical code coupled with chemical yields originating 
from SNe II, SNe Ia and single intermediate mass stars (IMS).
They concluded that a galactic wind develops as a 
consequence of the starburst and carries out of the galaxy mostly 
the metal-enriched gas and that the wind indeed is differential, in the 
sense that the elements produced by type Ia SNe are lost slightly more
efficiently than others. The reason for this resides in the fact
that when this kind of explosions (SNe Ia)
occurs, the medium is hotter and more rarefied than when the SNe II
occurs. In such a case, the efficiencies of energy 
transfer of supernovae type II and type Ia, contrary to 
BMD, are not the same. As the SNe Ia 
explosions occur in hotter a more rarefied medium, their 
energy is more efficiently thermalized into the ISM and, 
consequently, their efficiency of energy transfer is higher.

In this paper, in order to understand the chemical evolution 
histories of dSph galaxies and explore the possible connection 
between these galaxies and BCGs we use chemical evolution models 
which treat in details the energetics and take into account the 
yields of supernovae of type II and Ia as well as IMS. The goal
is to follow in detail the evolution of the abundance of several 
elements in both types of galaxies. The models are based on the 
work of BMD and follow the recent results 
for the energetics and winds given in Recchi et al. (2001). These 
prescriptions are very important since we assume that winds 
are the main responsible for the low gas content of the dSph 
galaxies together with the consumption of gas by star 
formation. The models for dSph galaxies and BCGs
differ in the prescriptions for star formation, in the luminous 
masses, in the wind efficiency and in the 
star formation efficiency.

The comparison of the models for both the dSph galaxies and BCGs
with the observational data of Damped Lyman $\alpha$ Systems (DLAs)
could also help to understand the nature of 
these systems,  since recent works using chemical and chemodynamical 
evolution models suggested that the progenitors of DLAs could be 
irregular galaxies such as the Large Magellanic cloud and/or BCGs 
(Calura, Matteucci $\&$ Vladilo 2003) and dwarf ellipticals 
(Lanfranchi $\&$ Fria\c ca 2003), respectively.

The paper is organized as follows: in Sect. 2 we present
the observational data concerning the dSph galaxies and BCGs,
in Sect. 3 the adopted chemical evolution models and star formation 
prescriptions are described, in Sect. 4 we describe the results of 
our models, which are compared to observational data from DLAs 
in Sect. 5 and finally in Sect. 6 we draw some conclusions.
We use the solar abundances measured by Grevesse $\&$ Sauval (1998)
when the chemical abundances are normalized to the solar values
([X/H] =  log(X/H) - log(X/H)$_{\odot}$) and a $H_0 = 70 \,km \, sec^{-1}\,
Mpc^{-1}$, $\Omega_m=0.3$, $\Omega_{\Lambda}=0.7$ cosmology is 
assumed throughout the paper.

\section{Data Sample}

The data sample collected for dSph galaxies consists of abundances 
of iron and some $\alpha$-elements such as O, Mg, Si and 
Ca obtained from the most recent high-resolution spectroscopy 
of red giant stars in 8 dSph galaxies of the Local Group 
(Smecker-Hane $\&$ McWilliam 1999; Bonifacio et al. 2000; 
Shetrone, Cot\'e $\&$ Sargent 2001; Shetrone et al. 2003). 
As the abundances were measured in stars within a range of ages 
it is possible to study the evolution with time of each galaxy 
or the dSph galaxies as a whole, if one assumes that these 
galaxies have a similar chemical evolution history. This is, 
though, a point of controversy, since there are suggestions 
in the literature that the local dSph galaxies exhibit complex star 
formation histories (Smecker-Hane $\&$ McWilliam 1999; Hernandez, 
Gilmore $\&$ Valls-Gabaud 2000). 
We will show that all these galaxies chemically evolve in a similar way, 
in spite of their different star formation histories, 
which are used as a constraint for the number, epoch and 
duration of the bursts of SF in the models. Besides the abundance 
ratios, we compare the predictions of the chemical evolution 
models to other properties of the dSph galaxies taken from the 
review of Mateo (1998).

For BCGs, the collected abundances of the elements N, O, C, Si 
and Fe were determined from measurements in HII regions
from either ground-based high resolution spectroscopy or 
from UV observations of the Hubble Space Telescope (Garnett et
al. 1995;  Kobulnicky $\&$ Skillman 1996; Izotov $\&$ Thuan 1999).
The diagram of the abundance ratios versus oxygen (e.g. N/O versus 
O/H) should not be viewed, in this case, as an evolutionary diagram 
because each point represents a different galaxy as it is 
seen at the present time, giving no clear information about their 
evolution (see Chiappini, Romano $\&$ Matteucci 2003). 
To compare properly these abundances with the prediction of 
the models one should only look at their final points, i.e the 
predicted present day abundances. However, as these galaxies might 
have been formed at different epochs, we plot also the evolutionary
lines in order to take into account possible different ages. 
It should be mentioned, though, that each point is a galaxy that 
could have followed a particular evolutionary track.

\section{Models} 

In order to study the chemical evolution and star formation 
histories of dSph galaxies and BCGs we use an updated version 
of the model of BMD in two different 
scenarios. The first one, representing the dSph galaxies, is 
characterized by a single and long star formation episode whereas
several short episodes of star formation separated by long 
quiescent periods are assumed for the BCGs.
These models allow one to follow in detail the evolution of 
the abundances of several elements, starting from the matter
reprocessed by the stars and restored into the ISM by winds and 
type II and Ia supernova explosions.

The main features of the models are the following:

\begin{itemize}

\item
one zone with instantaneous and complete mixing of gas inside
this zone;

\item
no instantaneous recycling approximation, i.e. the stellar 
lifetimes are taken into account;

\item
the evolution of several chemical elements (H, D, He, C, N, O, 
Mg, Si, S, Ca and Fe) is followed in detail;

\item
the efficiencies of energy transfer adopted are $\eta_{SNe II}$
= 0.03 , $\eta_{SNe Ia}$ = 1.0 and $\eta_{w}$ = 0.03 for
SNe II, SNe Ia and stellar winds, respectively (see BMD 
for more details);

\item
the nucleosynthesis prescriptions include the yields
of: Thielemann, Nomoto $\&$ Hashimoto (1996) for massive stars
(M $> 10 M_{\odot}$), van den Hoeck $\&$ Groenewegen 
(1997) for low and intermediate mass stars 
($0.8 \le M/M_{\odot} \le 8$) and Nomoto et al. 
(1997) for type I a SNe.
\end{itemize}

In our scenario, the dSph galaxies form as a result
of a continuous infall of pristine gas until a mass of
$\sim 10^8 M_{\odot}$ is accumulated.  One crucial 
feature in the evolution of these galaxies is the occurrence 
of galactic winds, which develop when the thermal 
energy of the gas equals its binding energy (Matteucci $\&$
Tornamb\'e 1987). This quantity is strongly influenced by 
assumptions concerning the presence and distribution 
of dark matter (Matteucci 1992). A diffuse ($R_e/R_d$=0.1, 
where $R_e$ is the effective radius of the galaxy and $R_d$ is 
the radius of the dark matter core) but massive 
($M_{dark}/M_{Lum}=10$) dark halo has been assumed for both models. 
The BCGs are built almost in the same way, but the total masses
eventually reached are higher ($\sim 10^{9} M_{\odot}$) as 
suggested by observations (these galaxies show large HI halos -
Mateo 1998). The occurrence of galactic winds in BCGs is not as 
important as for dSph galaxies. Actually, for some star formation 
efficiencies, winds in these systems do not even develop.

\subsection{Theoretical prescriptions} 

The time-evolution of the fractional mass of the element $i$ 
in the gas within a galaxy, $G_{i}$, is described by the basic 
equation:

\begin{equation}
\dot{G_{i}}=-\psi(t)X_{i}(t) + R_{i}(t) + (\dot{G_{i}})_{inf} -
(\dot{G_{i}})_{out}
\end{equation}

where $G_{i}(t)=M_{g}(t)X_{i}(t)/M_{tot}$ is the gas mass in 
the form of an element $i$ normalized to a total fixed mass 
$M_{tot}$ and $G(t)= M_{g}(t)/M_{tot}$ is the total fractional 
mass of gas present in the galaxy at the time t.
The quantity $X_{i}(t)=G_{i}(t)/G(t)$ represents the 
abundance by mass of an element $i$, with
the summation over all elements in the gas mixture being equal 
to unity. $\psi(t)$ is the fractional amount of gas turning into 
stars per unit time, namely the SFR. 
$R_{i}(t)$ represents the returned fraction of matter in the 
form of an element $i$ that the stars eject into the ISM through 
stellar winds and supernova explosions; this term contains all 
the prescriptions concerning the stellar yields and 
the supernova progenitor models. The two terms 
$(\dot{G_{i}})_{inf}$ and  $(\dot{G_{i}})_{out}$ account for 
the infall of external gas and for galactic winds, respectively.
The prescription adopted for the star formation history is the
main feature which characterizes a particular morphological 
galactic type.

The SFR $\psi(t)$ has a simple form and is given by:

\begin{equation}
\psi(t) = \nu G(t)
\end{equation}

The quantity $\nu$ is the efficiency of star formation, 
namely the inverse of the typical time-scale for star formation,
and is expressed in $Gyr^{-1}$.

In both cases, dSph galaxies and BCGs, $\nu$ is varied, but
in a larger range for dSph galaxies and with lower values than
the ones assumed for BCGs. The star formation proceeds even after 
the onset of the galactic wind but at a lower rate since a large
fraction of the gas ($\sim\, 10\%$) is carried out of the galaxy. 
In dSph galaxies the star formation is characterized by 1 or 2 
long episodes of star formation (3-13 Gyr), 
according to the star formation history of each individual galaxy as 
inferred by CMDs taken from Dolphin (2002) and Hernandez, Gilmore 
$\&$ Valls-Gabaud (2000). We also computed a model (the standard model)
for the whole  
sample of dSph galaxies using one long burst of star formation
starting at a galactic age of 1 Gyr with a duration of 8 Gyr. For 
BCGs, on the 
other hand, we adopted a SF proceeding in short bursts separated 
by long quiescent periods. The number of the bursts varies from 
2 to 7 and the duration of each burst from 10 to 200 Myr.

The rate of gas infall is defined as:
\begin{eqnarray}
(\dot G_{i})_{inf}\,=\,Ae^{-t/ \tau}
\end{eqnarray}

with A being a suitable constant and $\tau$ the infall timescale
which is assumed to be 0.5 Gyr.

The rate of gas loss via galactic winds for each element {\it i} is
assumed to be proportional to the star formation rate at the 
time {\it t}:

\begin{eqnarray}
 \dot G_{iw}\,=\,w_{i} \, \psi(t)
\end{eqnarray}

where $w_{i}$ is a 
free parameter describing the efficiency of the galactic
wind. The wind is assumed to be 
differential, i.e. some elements, in particular the products of 
SNe Ia, are lost more efficiently than others from the galaxy 
(Recchi et al. 2001). This fact translates into slightly different values 
for the $w_i$ corresponding to different elements. Here we will
always refer to the maximum value of $w_i$.
The efficiency of the wind is different
for each type of galaxy, being much higher for dSph galaxies than 
for BCGs.

The initial mass function (IMF) is usually assumed to be constant
in space and time in all the models and is expressed by the 
formula:

\begin{equation}
\phi(m) = \phi_{0} m^{-(1+x)}
\end{equation}

where $\phi_{0}$ is a normalization constant.
In both cases we assumed a Salpeter-like IMF (1955)
($x=1.35$), but tested also a flatter IMF ($x=1.1$)
and the Scalo (1986) prescription ($x=1.35$ for $0.1 \le 
m/m_{\odot} \le 2$, $x=1.7$ for $2 < m/m_{\odot} \le 100$),
always in the mass range $0.1-100 M_{\odot}$.

In table 1 we summarize the adopted parameters for the models
of BCGs and dSph galaxies.

\begin{table}
\begin{flushleft}
\caption[]{Model parameters for dSph galaxies and BCGs. $M_{tot}$ 
is the baryonic mass of the galaxy, $\nu$ is the star-formation 
efficiency and $w_i$ is the wind efficiency.}
\begin{tabular}{l|cccc}
\noalign{\smallskip}
\hline
\hline
\noalign{\smallskip}
 & $M_{tot} (M_{\odot})$&  $\nu(Gyr^{-1})$ &$w_i$  & $IMF$\\
\noalign{\smallskip}
\hline
\noalign{\smallskip}
dSph &$5*10^{8}$  &0.01-1 &10 &Salpeter\\
\noalign{\smallskip}
\hline
BCGs &$6*10^{9}$  &0.1-0.9 &0.5 &Salpeter\\
\hline
\end{tabular}
\end{flushleft}
\end{table}

\section{Results}

\subsection{Dwarf Spheroidal Galaxies} 

In order to understand the chemical evolution and star
formation histories of the dSph galaxies of the Local Group, we
compare the predictions of the chemical evolution model described
in the previous section with the observed [O/Fe], [Si/Fe], [Ca/Fe]
and [Mg/Fe] ratios. The final gas content and total mass of the 
modelled galaxy are also compared to the observational data.

Different models are computed by varying some parameters such as
the star formation efficiency, the number of the bursts, the 
duration of each burst, the wind efficiency, the total mass 
of the galaxy and the IMF prescription. The first choices for the
number and duration of the bursts were taken from the star 
formation histories inferred by CMDs of a sample of Local 
Group dSph galaxies and the star formation efficiency was fixed 
at low values as suggested by the observed abundance ratios (see Table 2).
The other parameters were tuned in order to match the observed 
quantities.

\begin{figure*}
\centering
\epsfig{file=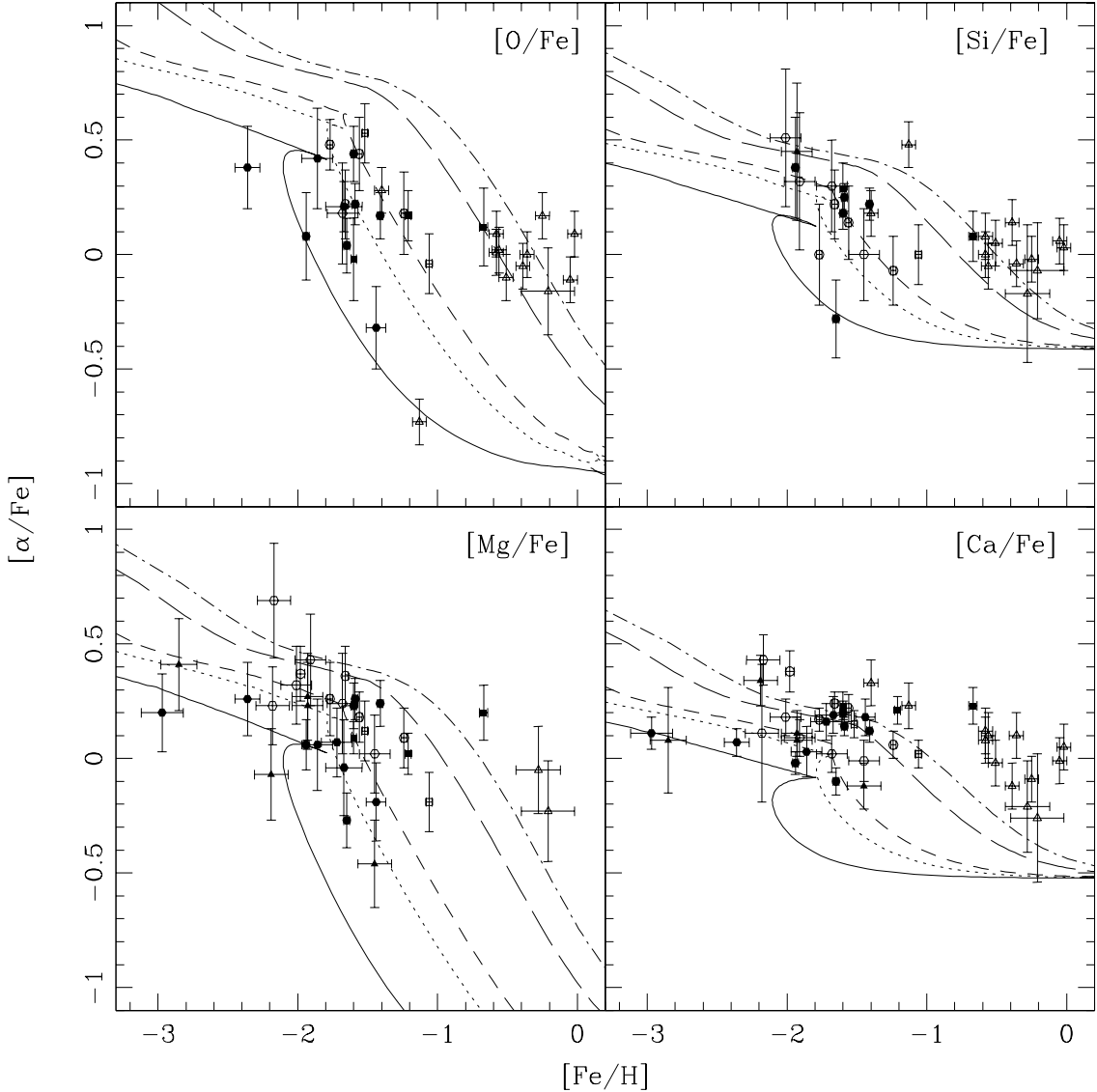,height=15cm,width=15cm}
\caption[]{[$\alpha$/Fe] vs. [Fe/H] observed in dSph galaxies 
compared to the predictions of the standard model for dSph galaxies.
The different lines correspond to different SF efficiencies 
(in $Gyr^{-1}$): 0.01 (solid line), 0.05 (dotted), 0.1 (dashed), 
0.5 (long-dashed), 1.0 (dotted-dashed). The signs represent stars 
of different 
galaxies: Sagittarius (open triangles), Draco (filled hexagons), 
Carina (filled circles), Ursa Minor (open hexagons), Sculptor
(open circles), Sextan (filled triangles), LeoI (open squares) and
Fornax (filled squares). The occurence of the wind is marked by 
the change in the slope of the curves.}
\end{figure*}

We started the study by considering the data sample as a whole,  
without separating individual galaxies. The similarity of 
the abundance patterns of the elements studied here for all 
the galaxies of the sample justifies this procedure and suggests 
that there is an uniformity in the chemical evolution 
of these galaxies, in spite of their different star formation
histories (Tolstoy et al. 2003).

One or two bursts with long duration (3 to 8 Gyr), a low 
efficiency for the star formation ($\nu < 1\, Gyr ^{-1}$) and
a high wind efficiency ($w_i > 5$) were chosen as 
initial parameters in order to reproduce the observed abundance 
ratios. We found that the most important parameter to reproduce 
the observed ratios is the star formation efficiency and the one
which best regulates the gas content, at a given SFR, 
is the wind efficiency. Three series of models with
different choices for the IMF were computed: one with a 
Salpeter-like IMF (1955), one with Scalo (1986) and a last one with
an exponent x=1.1. The models with a Scalo IMF fail in reproducing 
the observed abundance ratios, since they generally produce too 
low [$\alpha$/Fe] ratios for a given SF efficiency, thus were 
discarded. The other two 
models, with Salpeter and a flatter IMF, require both 
a star formation efficiency in the range $\nu$ $\sim$ 0.01 - 1 
$Gyr ^{-1}$ and a wind efficiency between $w_i \sim 5 - 15 
$ for a SF proceeding in one long episode of activity. 
The model with a flatter IMF, however, predicted too large
[$\alpha$/Fe] abundance ratios and HI mass/total mass ratios
so we adopted the Salpeter IMF. 
Figure 1 shows the observed abundance ratios compared with the 
predictions of the models with different efficiencies of star 
formation, which are characterized by one single episode of 
activity starting at 1 Gyr and lasting for 8 Gyr. The wind 
efficiency is $w_i$ = 10 and a Salpeter IMF is adopted
(from now on the standard model). The different
symbols in Figure 1 represent stars in different galaxies and the 
various lines correspond to different SF efficiencies.

\begin{table}
\begin{flushleft} 
\begin{center}\scriptsize  
\caption[]{Models for dSph galaxies. $n$, $t$ and $d$ are the number, time 
of occurrence
and duration of the bursts, respectively, and $\nu$ the range of
SF efficiencies.}
\begin{tabular}{lcccc}  
\hline\hline\noalign{\smallskip}  
model &n &t($Gyr$) &d($Gyr$) &$\nu$ ($Gyr^{-1}$)\\    
\noalign{\smallskip}  
\hline
standard model &1 &1 &8 &0.01-1 \\
Sextan  &1 &1 &8 &0.01-0.4\\
Sculptor &1 &0 &7 &0.03-0.5\\
Sagittarius &1 &0 &13 &0.5-3.0\\
Draco  &1 &6 &4 &0.005-0.05\\
Ursa Minor &1 &0.5 &3 &0.05-0.5\\
Carina &2 &6/10 &3/3 &0.01-0.5\\
\hline\hline
\end{tabular}
\end{center}
\end{flushleft}
\end{table} 
We noticed an interesting feature in the predicted abundance ratios: 
an abrupt decrease of [O/Fe] and [Mg/Fe] and a smoother one for
[Si/Fe] and [Ca/Fe] starting at [Fe/H] $\geq$ -2.0 for the model 
with $\nu=0.01\, Gyr ^{-1}$ and at [Fe/H] $\geq$ -1.0 for the one 
with $\nu =1\, Gyr ^{-1}$. These are the metallicities reached by 
the ISM when the wind develops. The behaviour shown in Figure 1 
has several reasons. The most 
important one is that after the onset of the wind the star formation
gradually vanishes and this decreases the [$\alpha$/Fe] ratios since 
the $\alpha$-elements are no more produced whereas Fe is produced by 
type Ia SNe.
In addition, we should consider the effect of the galactic wind:
as the efficiency of the wind is high 
and the wind is differential, a large amount of the processed gas is 
lost from the galaxy in different fractions for different elements
(D'Ercole $\&$ Brighenti 1999; McLow $\&$ Ferrara 1999).
When the SNe Ia explode the products of these explosions find an 
ISM hotter and more rarefied due to the 
effect of the previous explosions of SNII and are able to leave 
the galaxy more efficiently than the products of type II supernovae
(Recchi et al. 2001). As a consequence, the wind is more enriched 
in iron-peak elements, the principal products of SNe Ia,
than in $\alpha$ elements which are produced mainly in massive stars
and injected into the ISM by SNII explosions (Woosley $\&$ Weaver 
1995; Thielemann, Nomoto $\&$ Hashimoto 1996).
However, this effect is small in our formulation and it acts more 
on the absolute Fe abundance and on the models with lower star formation 
efficiency
(see for example the inversion of the [Fe/H]
in the model corresponding to $\nu=0.01Gyr^{-1}$). The decrease in the 
[$\alpha$/Fe] ratios is lower for Si and Ca than 
for O and Mg and the reason resides in the fact that Si and Ca are 
produced in type Ia SNe more than O and Mg.

The  behaviour of the [$\alpha$/Fe] ratios is consistent with the observed 
values which show
a clear decrease at metallicities higher than [Fe/H] $\sim$ -1.8
for [O/Fe] and [Mg/Fe] and a less intense decrease for [Si/Fe] and
[Ca/Fe] at the same metallicity. There are some points, 
however, which lie above the predictions of the models. Almost all
of them are stars from the Sagittarius dSph galaxy. This suggests
only that this galaxy requires a higher star formation efficiency 
than the values used in the standard model (see Figure 2). From 
{Figure 1 one can see also that the wind is not important only in 
regulating the remaining gas content, but also in determining the 
pattern of abundance ratios, since the abundance ratios observed 
in many of the stars of the local dSph galaxies are reproduced 
only with the occurrence of an intense wind. In particular, 
the effect of the wind is to decrease the absolute Fe abundance while 
the [$\alpha$/Fe] ratios
are not so affected since both $\alpha$-elements and Fe are lost. 
The fact that Fe is lost with a sligthly higher efficiency than the 
$\alpha$-elements is not the dominant factor in these plots.
We have run also models with a normal galactic wind (same efficiency for 
all the elements) and the difference in the resulting curves is quite 
small but the effect is more evident in the model with the lowest star 
formation efficiency.
The range adopted for the star formation efficiency allows the 
standard model to reproduce the data of all the studied galaxies, 
assuming that they have similar star formation histories, in 
particular
one long starburst. The differences in the abundance ratios
in different galaxies are then mainly a combined effect of the wind 
and star formation efficiencies rather than due to the detailed
star formation histories.
In order to check this assumption we computed different models 
for each galaxy using the star formation histories inferred by
CMDs (Hernandez, Gilmore $\&$ Valls-Gabaud 2000; Dolphin 2002) 
as a constraint on the number, epoch and duration of the bursts (see Table 2). 
The efficiency of the star formation was varied to get the best 
agreement with all the $\alpha$/Fe ratios observed in each galaxy. 
The other parameters as the efficiency of the wind, 
IMF, initial mass, amount and distribution of dark matter, were 
the same as in the standard model. We used the 6 galaxies for which 
there are enough data to compare with the predictions of the models,
in particular Draco, Ursa Minor, Sculptor, Sagittarius and Carina, 
for which a star formation history has been inferred
(see table 2 for more details). For the last one, Sextan, we 
used the prescriptions of the standard model. In Figure 2 we 
show the predictions of the individual models with 
different SF efficiencies compared with the [Si/Fe] (or [Mg/Fe] when Si 
is not available) observed in each galaxy. It should be mentioned 
that the SF efficiencies determined for each galaxy were inferred 
from the comparison with the four abundance ratios observed, 
even though only the data for one ratio are shown in Figure 2.

\begin{figure}
\centering
\epsfig{file=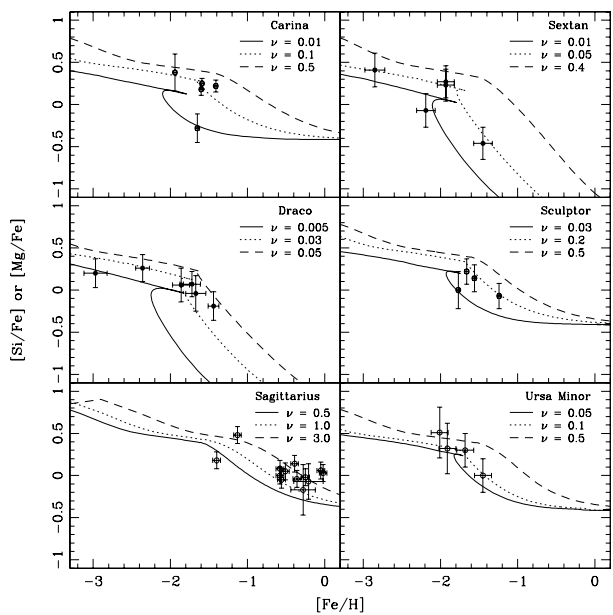,height=8cm,width=8cm}
\caption[]{Predictions of the models for individual dSph galaxies 
compared to observed [Si/Fe] (open circles) or [Mg/Fe] (filled 
circles). The lines represent different SF efficiencies.}
\end{figure}

When using the number, duration and epoch of the bursts given
by the inferred star formation histories, coupled with the
star formation efficiencies given by the standard model, the data
of all galaxies are well reproduced.
This suggests that, indeed, the most important parameter
to explain the observed distributions is the efficiency 
of the star formation. One can see, from Figure 2, that 4 
(Carina, Sextan, Sculptor and Ursa Minor) out of the 6 galaxies 
require a similar range of SF efficiencies, $\nu$ = 
0.01-0.5 $Gyr^{-1}$. The other 2 galaxies are reproduced by either
a lower (Draco) or a higher  $\nu$ (Sagittarius). 

\begin{figure}
\centering
\epsfig{file=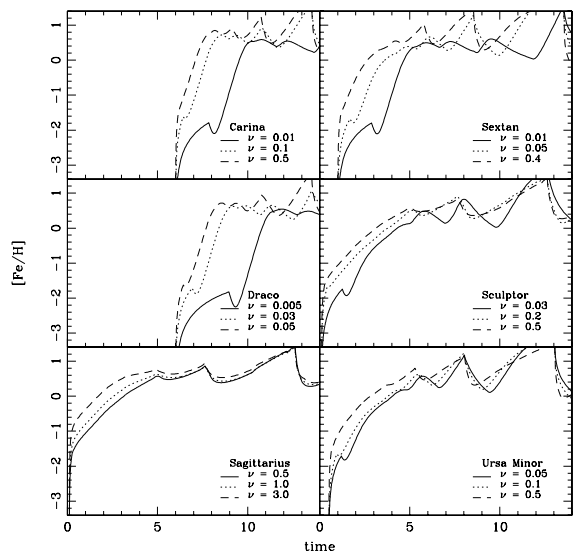,height=8cm,width=8cm}
\caption[]{[Fe/H] as a function of time as predicted by the models
for each individual dSph galaxy. The different lines represent different 
SF efficiencies.}
\end{figure}

The differences in the chemical evolution among these galaxies
can be clearly seen in Figure 3, where the predicted Fe-age relations are 
plotted for each individual dSph galaxy.
At variance with the abundance ratios, [Fe/H] is more sensitivy to
the details of the SF history of the galaxy as one can see in Figure 3.
The galaxies could be divided in two groups: one including four galaxies
(Sagittarius, Ursa minor, Sculptor and Sextan), where the SF starts at
early times, and another one with the 
remanining two galaxies (Draco and Carina), where the SF starts at an
older cosmic time. In the galaxies of the second group the [Fe/H] reach 
the solar value faster ($\sim$ 1 - 2 Gyr after the onset of the SF) than in 
the galaxies of the first group ($\sim$ 2 - 5 Gyr) and there is a narrower 
range of ages for the stars which are younger
(Draco and Carina) if compared to the other
galaxies at the same metallicity. Consequently, we predict that one should
observe a wide spread in ages 
of the stars of Sagittarius, Sculptor, Ursa Minor and Sextan (a little 
bit lower in this latter) whereas the observed stars of Carina and
Draco should have ages within a short interval of some Gyrs. It should be 
mentioned, though, that this last statement depends on the adopetd SF 
histories. Another point that should be viewed with care is the 
Fe abundances which reach high values, well above the ones observed in 
stars of the dSph galaxies of the local group. The predicted
[Fe/H] is, in fact, the abundance expected in the ISM, and as the SF 
gradually vanishes and the galactic wind develops, the number of stars 
formed after this
stage is very low. This stage corresponds to a metallicity range of 
[Fe/H] $\sim$ -1.5 - -0.5 dex, depending on the SF efficiency, and
the number of stars formed with metallicities higher than these
will decrease as the metallicity in the ISM increases.

\begin{table*} 
\begin{center}\scriptsize  
\caption[]{Predictions of the models for dSph galaxies compared to 
observations. M$_{tot}$ is the present day total mass of the galaxy,
 M$_{HI}$/M$_{tot}$ is the present day ratio between the HI mass and
total mass.}
\begin{tabular}{c|cc|cc|cc|cc}  
\hline\hline\noalign{\smallskip}  
   &\multicolumn{2}{|c|}{M$_{tot}$ (10$^6$ M$_o$)}  
&\multicolumn{2}{|c|}{M$_{HI}$/M$_{tot}$}  
&\multicolumn{2}{|c|}{1$^o$burst (Gyr)} 
&\multicolumn{2}{|c|}{2$^o$burst (Gyr)}\\    
\noalign{\smallskip}  
\hline
 &obs &mod &obs &mod &obs &mod &obs &mod\\
\hline
standard model &6.4-68$^d$ &6.4-85.2 &$<$0.004$^d$ &0.0001-0.0004 
&- &1-9 &- &- \\
Sextan &19$^a$ &6.4-46.4 &$<$0.001$^a$ &0.0001-0.0004 &- &1-9  &- &-\\
Sculptor  &6.4$^a$ &9.6-40.1 &0.004$^a$ &0.0002-0.0003  &0-7$^c$ &0-7 &- &-\\
Sagittarius &-- &40.1-150.7 &$<$0.001$^a$ &0.0002  &0-13$^c$ &0-13 &- &-\\
Draco  &22$^a$ &5-21.1  &$<$0.001$^a$ &0.0002-0.0007  &6-10$^b$ &6-10 &- &-\\
Ursa Minor  &23$^a$  &12.5-47.6 &$<$0.002$^a$ &0.0002-0.0004  &0-3$^b$ &0.5-3.5
&- &-\\
Carina &13$^a$  &6.6-56.5 &$<$0.001$^a$ &0.0004-0.0006 &6-9$^b$ &6-9 &10-13$^b$ 
&10-13\\
\hline\hline
\end{tabular}

a - Mateo (1998)

b - Hernandez, Gilmore $\&$ Valls-Gabaud (2000)

c - Dolphin (2002)

d - for the whole sample of dSph galaxies in Mateo (1998)
\end{center}
\end{table*} 
\begin{figure}
\centering
\epsfig{file=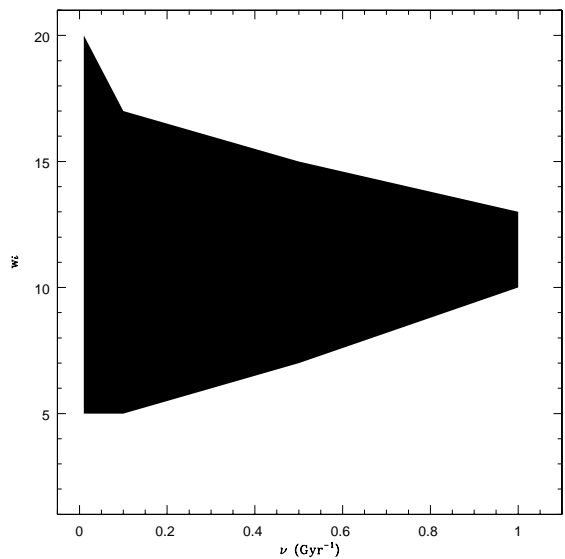,height=8cm,width=8cm}
\caption[]{Values of the wind efficiency which fit the data of the
dSph galaxies as a function of the SF efficiency. The shaded area 
correspond to the values of both parameters which reproduce the observed
data.}
\end{figure}

The adopted wind efficiency is the same for all galaxies, even 
though differences in the rate of wind might exist. 
These differences in the wind rate between galaxies
are accounted for by the various SF efficiencies, since 
the rate of the wind is assumed to be proportional to the SFR.
As the SF efficiencies are similar for the galaxies of the studied
sample, but not equal, so it is the wind rate. The wind efficiency
establishes how intense is the wind, i.e. the amount of material 
which is actually lost from the galaxy. The results of our model for the 
present day total mass and gas content of the galaxies show that 
the range of wind efficiencies adopted here provides a good 
fit to the observed data (see Table 3), but it can vary depending 
on the adopted SF efficiency (Figure 4). If the wind efficiency
is lowered below $w_i=5$, the final gas mass in 
the modelled galaxies with high SF efficiencies ($\nu \geq 0.1 
\, Gyr^{-1}$) would be much higher than the values observed and the
decrease in the abundance ratios of the models with low SF efficiencies 
($\nu \leq 0.1 \, Gyr^{-1}$) would not be enough to reproduce the lowest 
values observed. If, on the other side, $w_i$ were too high 
($w_i \geq 20$ ) the galaxy, in the models with the lowest 
SF efficiencies ($\nu \leq 0.1 \, Gyr^{-1}$), would loose all the gas 
content as soon as the wind starts, the star formation would 
cease and there would not be stars forming with metallicities higher 
than that in the gas when the wind develops ([Fe/H]$\sim$ -2.0 for the 
model with $\nu=0.01\,Gyr^{-1}$ and [Fe/H]$\sim$ -1.0 for the model with  
$\nu=1\,Gyr^{-1}$). On the other hand, in the models with high SF 
efficiencies a high wind efficiency ($w_i \geq 13 $) 
predicts abundance ratios which would not reproduce the highest values 
observed. Given these facts and Figure 4, one can conclude that the 
range of the wind efficiency becomes narrower as the SF efficiencies 
increases and that a value between $w_i \sim 10 - 13$ can 
reproduce the observed final masses and abundance ratios independently
from the adopted SF efficiency.

\begin{figure}
\centering
\epsfig{file=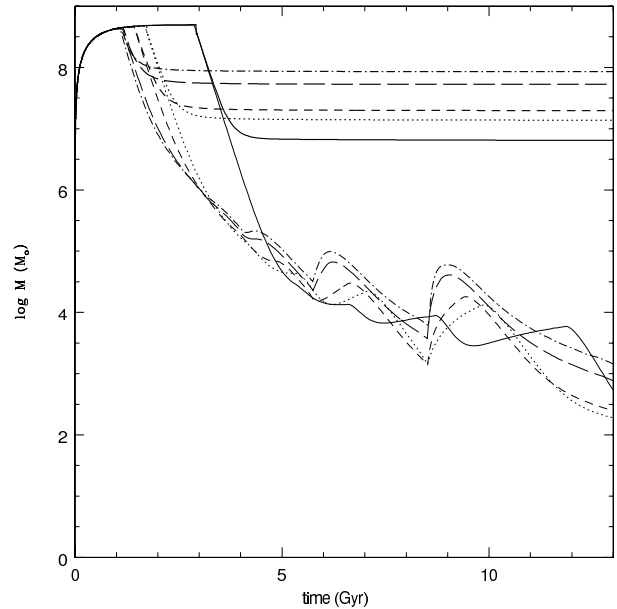,height=8cm,width=8cm}
\caption[]{Evolution with time of the total mass (thick lines) and
gas content (thin lines) as predicted by the standard model for 
dSph galaxies with different SF efficiencies.
The different lines correspond to different SF efficiencies 
(in $Gyr^{-1}$): 0.01 (solid line), 0.05 (dotted), 0.1 (dashed), 
0.5 (long-dashed), 1.0 (dotted-dashed). }
\end{figure}

In Figure 5 we plot the evolution with time of the total mass
(thick lines) of the galaxy and its gas content (thin lines)
for the standard model with different efficiencies of star 
formation (the different lines). The time of the occurrence
of the wind varies with the rate of star formation since it 
depends on the amount of energy which is injected into the ISM. 
For the model where the star formation is less efficient (solid 
line), a wind is developed only almost 2 Gyr after the onset of 
the star formation, but for the cases where the star formation is
more intense the wind occurs sooner ($\sim$ 100 Myr) after the star
formation begins. As the wind is very efficient, the gas content 
and total mass of the galaxy decrease fast in few Gyrs, but after 
that, the total mass remains almost constant and the amount of gas 
continues to decrease at a much lower rate, in agreement with 
Ferrara $\&$ Tolstoy (2000). The oscillation in the gas content 
is due to the duration of the wind and to the contemporary 
injection of gas into the ISM from dying stars. 
The results for the total final mass and gas content of
the standard model applied to each individual galaxy and  compared to 
observational data are shown in Table 3. It is also shown the 
number and epochs of the bursts of star formation. The models for 
each galaxy as well as the standard model are in good agreement 
with the observational data. The variations of the masses in the
models are due to the range of star formation efficiencies adopted
in each model. In general, the larger the star formation 
efficiency, the higher the total final mass and the lower the final
gas mass/total mass ratio, since more stars are formed and 
consequently more gas is consumed (see also Figure 5).

\subsection{Blue Compact Galaxies}

The possible scenarios for BCGs proposed in the literature in order 
to explain 
the pattern of several abundance ratios consider these galaxies 
either as young objects, especially those with 12 + log (O/H) 
$\leq$ 7.6 (Izotov $\&$ Thuan 1999), where primary N produced 
in massive stars is required to explain the low values of 
N/O, or older systems which suffered several episodes of star 
formation separated by long quiescent periods, in which the low 
values of N/O are a consequence of the time-delay between
the productions of O in massive stars and N in IMS 
(Marconi, Matteucci $\&$ Tosi 1994; Garnett et al. 1997; 
BMD; Larsen, Sommer-Larsen $\&$ Pagel 2001). In order to explore 
these two possibilities we made use of 
different galaxy models following the prescriptions and results of
BMD and Recchi et al. (2001, 2002), as discussed previously.
Other parameters such as 
the number and duration of the bursts, the efficiency of the star 
formation and the galactic wind, the slope of the IMF, the 
production of N, regarding its primary or secondary origin 
in massive stars, were varied in order to understand the
observed distribution of N/O, C/O, Si/O and [O/Fe] versus
O/H in BCGs.

First we defined a standard model fixing some parameters:
a Salpeter IMF, a star formation efficiency between $\nu$ = 0.05 - 2 
$Gyr^{-1}$}, a wind efficiency $w_i$ = 0.5,
a ratio of dark to luminous matter of 10
and only secondary production of N in massive stars. The 
number and duration of the bursts were then varied in 
order to get the best fit to the observed distributions.
We computed models with 2 bursts and durations between
10 Myr to 1 Gyr and increased the number of bursts up to
7, with duration no longer
than 0.2 Gyr each. As the number of the bursts increases, the duration
of each burst decreases, otherwise the galaxy is disrupted by 
galactic winds.

\begin{figure}
\centering
\epsfig{file=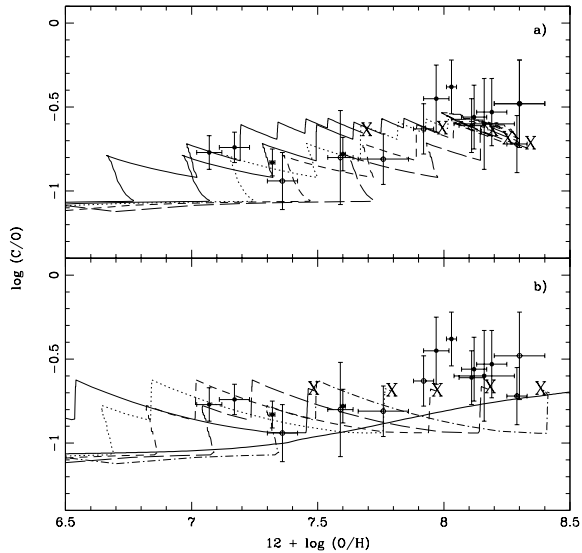,height=8cm,width=8cm}
\caption[]{log (C/O) vs. 12 + log (O/H) observed in BCGs compared
to the predictions of the standard model for BCGs with 7 and 4 
bursts (panels $a$ and $b$, respectively). The lines represent 
different SF efficiencies (in $Gyr^{-1}$): 0.1 (solid line), 
0.2 (dotted), 0.3 (dashed), 0.5 (long dashed) and 0.9 (dotted-dashed). 
The solid thick line represents a model with continuos
star formation and a SF efficiency $\nu$ = 0.1  $Gyr^{-1}$.
The crosses represent the final abundances
of the models. Filled  symbols are the data from Izotov $\&$ Thuan
(1999) and open symbols from Garnett et al. (1995).}
\end{figure}
\begin{figure}
\centering
\epsfig{file=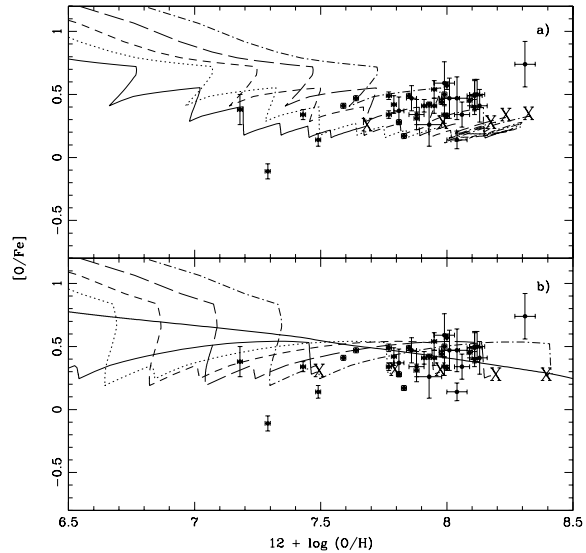,height=8cm,width=8cm}
\caption[]{[O/Fe] vs. 12 + log (O/H) observed in BCGs compared
to the predictions of the standard model for BCGs with 7 and 4
bursts (panels $a$ and $b$, respectively). The solid thick line 
represents a model with continuos
star formation and a SF efficiency $\nu$ = 0.1  $Gyr^{-1}$.
The symbols are the 
same as in Figure 6. The data are from Izotov $\&$ Thuan
(1999).}
\end{figure}

After defining a range for the number of the bursts and their
duration we varied also other parameters such as the exponent of
the IMF, the star formation and wind efficiencies, and the
origin of N in massive stars (primary and/or secondary). As already
noticed by BMD, the most important parameter to reproduce
the observed N/O, C/O, Si/O and [O/Fe] versus O/H distributions is 
the star formation efficiency. We found that a star formation
efficiency in the range $\nu$ = 0.1-0.9 $Gyr^{-1}$ is needed in 
order to reproduce the majority of the observational constraints
when a Salpeter IMF is adopted. The range of values 
for $\nu$ changes when a Scalo IMF is used but remains the 
same if an exponent x=1.1 is adopted. In the case of a Scalo IMF,
the values of the star formation efficiency required to reproduce
the observed metallicities must be higher 
but the agreement with the observed abundance ratios is poorer. 
An IMF with an exponent x=1.1 should also be discarded. In this 
case, in fact, the number of massive stars produced is
higher and so the abundance of $\alpha$ elements, the main
product of these stars. Consequently, the N/O ratio is well 
reproduced, especially the lower values, but at the same time, 
the predicted [O/Fe] ratios become too high and the C/O ratios too 
low if compared with the observations.

In the range of the star formation efficiency defined above for
a Salpeter IMF, models with 2 to 7 bursts are able to reproduce the
observed abundance ratio diagrams if proper values
for the wind efficiency and durations of the bursts are adopted.
The difference between these models is that as the number of bursts
increases, increases also the variation in the predicted ratios 
(sawtooth behaviour) and the spread of the data is better 
reproduced by the models with 7 bursts (standard model $a$ in 
table 4) in the case of C/O,
[O/Fe] and Si/O (panel $a$ in Figures 6, 7 and 8, respectively). 
The N/O, especially the lowest values, are however difficult to 
be achieved with 7 bursts (panel $a$ in Figure 9), without 
breaking the agreement with the other ratios.

\begin{figure}
\centering
\epsfig{file=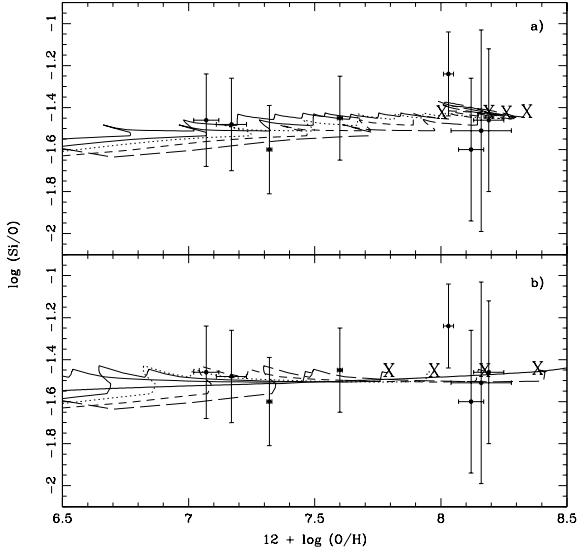,height=8cm,width=8cm}
\caption[]{log (Si/O) vs. 12 + log (O/H) observed in BCGs compared
to the predictions of the standard model for BCGs with 7 and 
4 bursts (panels $a$ and $b$, respectively). The solid thick line 
represents a model with continuos
star formation and a SF efficiency $\nu$ = 0.1  $Gyr^{-1}$.
The symbols are the 
same as in Figure 6. The data are from Izotov $\&$ 
Thuan (1999).}
\end{figure}

\begin{figure}
\centering
\epsfig{file=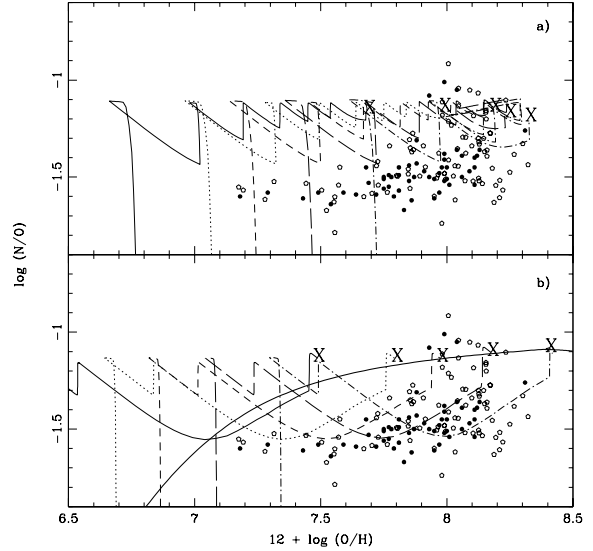,height=8cm,width=8cm}
\caption[]{log (N/O) vs. 12 + log (O/H) observed in BCGs compared
to the predictions of the standard model for BCGs with 7 and 
4 bursts (panels $a$ and $b$, respectively). The solid thick line 
represents a model with continuos
star formation and a SF efficiency $\nu$ = 0.1  $Gyr^{-1}$.
The symbols are the 
same as in Figure 6. Filled  symbols are the data from Izotov $\&$
Thuan (1999) and open symbols from Kobulnicky $\&$ Skillman (1996).}
\end{figure}

A model with 4 bursts (standard model $b$ in table 4), instead, 
reproduces all the values of N/O, including the low ones, and 
fits the other ratios as well (panel $b$ in Figures 6, 7, 8 and 9)
if the epochs and durations of the bursts are 
properly chosen. A first short burst occurring at 1 Gyr with 
a duration of 20 Myr and three more recent bursts at 11, 13 and 
13.98 Gyr lasting for 10, 200 and 20 Myr, respectively, produce the 
best fit to the data. The low values of N/O are achieved during  the 
third burst when the value of this ratio decreases due 
to the production and injection of O in the ISM by massive stars. 
The low values for N/O at low O/H, in this case, are not 
characteristic of very
young systems as suggested in the literature (Izotov $\&$ 
Thuan 1999) since it is necessary to have at least two previous bursts
in order to reproduce these values. Actually, these are old 
systems with several stellar populations formed in different bursts
at different epochs, as suggested by the colours of BCGs indicating
an underlying old population (Papaderos et al. 1996b; Doublier et
al. 1997, 2000; Marlowe, Meurer $\&$ Heckman 1999).
\begin{table*}
\begin{center}\scriptsize  
\caption[]{Models for BCGs. $n$, $t$ and $d$ are the number, time 
and duration of the bursts, respectively, and $\nu$ the range of
SF efficiencies.}
\begin{tabular}{lcccc}  
\hline\hline\noalign{\smallskip}  
model &n &t($Gyr$) &d($Gyr$) &$\nu$ ($Gyr^{-1}$)\\    
\noalign{\smallskip}  
\hline
standard model a &7 &0.5/1/2/5/7/11/13.9 
&0.05/0.05/0.05/0.05/0.05/0.05/0.1 &0.1-0.9\\
standard model b &4 &1/10/13/13.98 &0.02/0.01/0.2/0.02
&0.1-0.9\\

\hline\hline
\end{tabular}
\end{center}
\end{table*} 

At this point it should be mentioned that the predicted final
abundance ratios of all the models (the crosses in the figures)
are in agreement with the observed ratios, except for the model 
with 4 bursts and $\nu=0.9\, Gyr^{-1}$ (dotted-dashed line in panel
b in Figures 6, 7, 8 and 9). This model, in fact, indicates that 
systems with higher SFR should be 
younger in order to fit the data. In particular, the evolutionary
lines suggest that: younger systems (no more than 1 Gyr) with
the same evolutionary history can also be representative of the
population of BCGs. This is especially true for the galaxies with
low values of N/O at low O/H, which are reproduced by the models
during their third burst of SF which occurred 1 Gyr before
the present day burst. The almost constant values of N/O ($\sim$ 
-1.6 dex) of these systems could, therefore, be representative of
an isochrone since all of them are reached at a galatic age of 
13 Gyr.

We also runned models with continuos star formation (thick solid
line in Figures 6, 7, 8 and 9) in the same range of SF efficiencies 
used in the bursting models (0.1 - 0.9 $Gyr^{-1}$). In this case, the 
predictions of the models can hardly reproduce at the same time 
the O/Fe, C/O, Si/O and N/O ratios during
the evolution of the galaxy and predict a too high present day O/H
(which lies out of the range of the plots) compared to the ones observed
in the majority of the BCGs. Besides that, the amount of gas compared to 
the mass of stars is lower than the ones inferred for this type of galaxy.
One can conclude, therefore, that a continuous star formation may not
be a representative scenario for the population of BCGs.

\begin{figure}
\centering
\epsfig{file=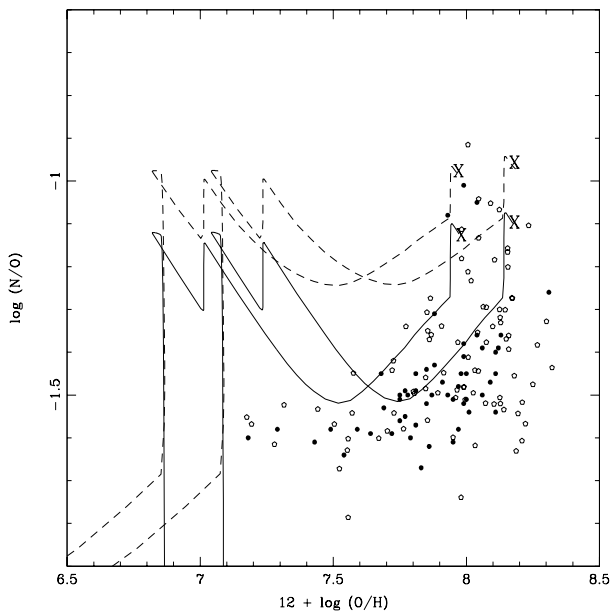,height=8cm,width=8cm}
\caption[]{log (N/O) vs. 12 + log (O/H) observed in BCGs compared
to the predictions of the standard model for BCGs with 4
bursts and $\nu$ = 0.3 and 0.5 $Gyr^{-1}$ assuming primary 
production of N in massive stars in different amounts (solid 
and dashed lines). Filled  symbols are the data from Izotov $\&$
Thuan (1999) and open symbols from Kobulnicky $\&$ Skillman (1996).
The crosses represent the final points of the models.}
\end{figure}

It is worth noticing that the production of primary N in massive
stars is not required in 
order to reproduce the observed N/O, as already noticed by 
Chiappini, Romano $\&$ Matteucci (2003), who studied the origin 
of carbon, oxygen and nitrogen using chemical evolution models 
for our Galaxy, more massive disk galaxies and irregular galaxies.
In Figure 10 are shown the predictions of the models with 4 bursts
when different amounts of primary N produced by massive stars are
assumed. The prescriptions for the production of primary nitrogen 
are the same as in Matteucci (1986). In fact, when 
large amounts of primary N produced by massive stars are adopted 
in our models, the predicted N/O ratios are increased and the 
evolutionary lines grow too fast thus producing a
result at variance with the data (dashed lines in Figure 10). 
Only if very low amounts of N are produced by massive stars, 
roughly in agreement with the yields predicted by models of 
stellar evolution including rotation (Meynet $\&$ Maeder 2002),
this effect would  hardly be noticed and the predictions of the 
models would still reproduce the observed distributions (solid 
lines in Figure 10).

\subsection{Connection between dSphs and BCGs}

Dwarf spheroidal galaxies are evolved systems, as witnessed 
by the lack of detectable ISM.
The 
evolution in these systems could occur in two ways: one type of dwarf galaxy 
could 
evolve to another type due to internal (SF, winds) or external 
(mergers, interactions) effects. Alternatively, the type of the galaxy was 
determined by the initial conditions and no further change in 
the morphology has occurred since then.
Here, we focus on the possible connection between different 
types of dwarf galaxies in an unified evolutionary scenario.

\begin{figure}
\centering
\epsfig{file=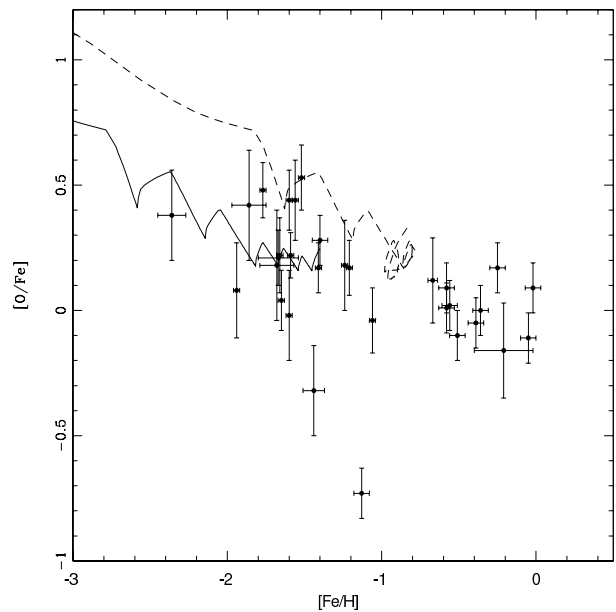,height=8cm,width=8cm}
\caption[]{[O/Fe] vs. [Fe/H] observed in dSph galaxies compared
to the predictions of the standard model for BCGs with 7 bursts.
The solid line represents a model with a SF efficiency $\nu
=0.1\ Gyr^{-1}$ and the dashed line represents a model with 
$\nu=0.9\ Gyr^{-1}$. No model for dSph galaxies can fit the 
data for the BCGs.}
\end{figure}

Among the scenarios proposed so far, the one which connects 
BCGs and dSph galaxies could be investigated through the 
chemical evolution
models described in the previous sections. In this scenario
a starburst in a BCG gives rise to a super wind which removes
all the gas of the galaxy and halts the SF, giving rise to
a dSph galaxy. We see, however, that the chemical evolution 
histories inferred by the models which give the best match to 
the observational data of BCGs and local dSph galaxies exhibit
marked differences. In particular, the 
major differences
are related just to the SF history and to the galactic wind.
While the observed distributions of N/O, C/O, Si/O and [O/Fe]
in BCGs are reproduced by models with SF characterized by short
and repeated bursts separated by long quiescent periods, the 
abundance ratios of local dSph galaxies require a SF which 
proceeds 
in one (or two) long episodes of activity. The SF efficiencies 
of the two models also differ even though they overlap in the 
highest values. The similarity between the SF efficiencies, 
however, could raise the question whether a bursting
SF could reproduce the dSph data or the distribution of abundance 
ratios in BCGs could be fitted by models characterized by one
or two long bursts. The latter possibility is very difficult to
achieve, if not improbable (thick solid lines in Figures 6, 7, 8
and 9), but the former is more likely and was 
already suggested for some dSph galaxies such as Carina. 
In this case, however, the very low values of $\alpha$/Fe at 
high metallicities, especially O and Mg, are not reproduced 
by this kind of model, as one can see in Figure 11, where we show
the observed [O/Fe] ratio in dSph galaxies compared to the predictions
of the model applied to BCGs.

The lack of agreement between the models with several bursts and
a low wind efficiency and the data of dSph galaxies clarifies the
different roles played by the wind and the SF in the evolution of 
BCGs and dSph galaxies. As already seen in section 4.1, the 
occurrence of a wind is a crucial feature not only in the evolution 
of the total baryonic mass and gas content, but also in the 
chemical enrichment history of the dSph galaxies: a high wind 
efficiency ($w_i= 5-15$) is required in order to match the 
observed abundance ratios and stellar masses. On the other hand, 
the distribution of abundance ratios in BCGs are reproduced by a 
model with a low wind efficiency, $w_i=0.5$. Little  
variations in this value would not change the results as long as
$w_i\, <\, 1$. If the wind efficiency is larger than 
$w_i\, =\, 1$ the variations in the predicted abundance
ratios would be much more intense and would produce a poorer fit to 
the data. 
Besides that, the final gas masses of these galxies would be
well below the observed ones. These low values mean that the occurrence
of the galatic winds in BCGs does not influence significantly
their evolution. In fact, in the models with the lowest SFR,
galactic winds do not even develop. Besides that, a model with 
short episodes of SF separated by long quiescent periods within
the range of SF efficiency adopted here is not able to produce 
the relatively high oxygen abundances observed in dSph galaxies
in comparison with the ones observed in BCGs.

From the above remarks, it seems unlikely, from the chemical 
evolution point of view, that the BCGs evolve into dSph galaxies
after a superwind triggered by several bursts of SF. A more
reliable scenario would be the one in which the initial conditions
determined the future evolution of these galaxies as proposed by 
Ferrara $\&$ Tolstoy (2000). These authors suggested that
some dwarf galaxies might have started their evolution from
the same progenitors, but would have followed different
evolutionary tracks and that BCGs represent simply  
a normal tail of the distribution of dwarf galaxies characterized
by higher central mass density which favors the onset of 
occasional starbursts. Some dSph galaxies, instead, could have 
been formed from protogalactic objects (Pop. III objects - 
Couchman $\&$ Ress 1986, Ferrara 1998, Ciardi, Ferrara $\&$ Abel 
1999) at very high redshifts ($z_f\,\sim\,15$) and lost their gas 
content either by blowaway or photoionization.

\section{Damped Lyman $\alpha$ Systems}

The Damped Lyman $\alpha$ Systems are the QSO absorption line 
systems with the highest neutral hydrogen column densities (N(HI) 
$\geq \, 10^{20}\,cm^{-2}$). These systems are believed to be the 
progenitors of the present day galaxies for several reasons: they 
dominate the neutral gas content of the universe (Wolfe et al. 1995; 
Rao $\&$ Turnshek 2000), and the comoving mass density of gas inferred 
from DLAs at $z\, =\, 2-3$ is comparable to the current baryonic 
mass density of stars at the present time (Wolfe et al. 1995; 
Storrie-Lombardi $\&$ Wolfe 2000). As DLAs can be observed at 
very high redshifts (up to $z \sim 4.4$) they can probe the very early
stages of evolution of galactic systems. Thus, the understanding 
of their nature is a key factor in our 
knowledge of the formation and evolution of galaxies.
  
The nature of DLAs at high redshifts, however, is still a matter of 
debate and several hypothesis have been drawn recently, including
disk galaxies, dwarf irregular galaxies, low surface brightness 
galaxies (Wolfe et al. 1986; Prochaska $\&$ Wolfe 1997, 1998; 
Pettini et al. 1999; Centurion et al. 2000; Boissier $\&$ 
Prantzos 2000; Molaro et al. 2001). A picture of
a multi-population of galaxies being responsible for the absorption 
seen in DLAs is emerging from studies of possible hosts of 
DLAs which revealed a wide range of morphological types and 
luminosities (Lanzetta et al. 1997; Pettini et al. 2000;
Turnshek et al. 2001).

Several studies addressing the question of the nature of DLAs
from the chemical evolution point of view appeared in the last 
few years and some connections between the DLAs and dwarf galaxies
were suggested by studies using chemical evolution or chemodynamical
models compared to a series of abundance ratios observed in DLAs.
In a germinal paper Matteucci, Molaro $\&$ Vladilo (1997) compared 
the predictions of chemical evolution models with observed 
$\alpha$/Fe and N/$\alpha$ ratios and suggested that starburst 
galaxies could be the progenitors 
of some DLAs. More recently, Calura, Matteucci $\&$ Vladilo
(2003), using chemical evolution models representing galaxies of 
different morphological types, pointed out that the majority of 
DLAs can be explained either by disks of spirals observed at large 
galactocentric distances or by starburst dwarf irregular galaxies
and irregular galaxies such as the Large Magellanic Cloud. The 
dwarf spheroids were also claimed as possible progenitors 
for the DLAs by Lanfranchi $\&$ Fria\c ca (2003) who used a 
chemodynamical model. In their simulations, a series of
starbursts driven by gas flows inside the galaxy are able 
to reproduce the majority of the DLAs. In this work, we 
considered also the BCGs and the dSph galaxies as possible 
candidates in our attempt to recover the nature of DLAs. The 
models used are the same as described in the previous sections.
For the models of BCGs, though, we used also higher SF efficiencies
relative to the standard model in order to account for all the
observed systems. 

\begin{figure}
\centering
\epsfig{file=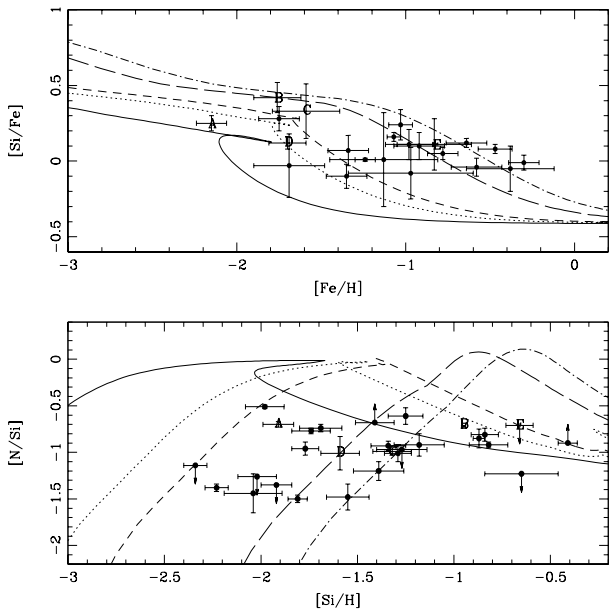,height=8cm,width=8cm}
\caption[]{[Si/Fe] vs [Fe/H] corrected for dust (above) and [N/Si]
 vs [Si/H] (below) observed in DLAs compared to the predictions of
the standard model for dSph galaxies with different SF efficiencies
(in $Gyr^{-1}$):
 0.01 (solid line), 0.05 (dotted), 0.1 (dashed), 0.5 (long-dashed),
 1.0 (dotted-dashed). The letters correspond to DLAs with both 
ratios observed.}
\end{figure}

\begin{figure}
\centering
\epsfig{file=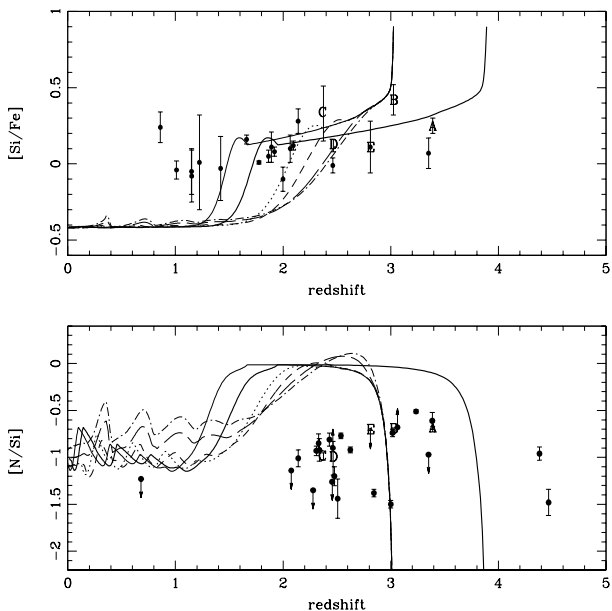,height=8cm,width=8cm}
\caption[]{Evolution with redshift of [Si/Fe] corrected for dust (above)
and [N/Si] (below) observed in DLAs compared to the predictions of the
standard model for dSph galaxies with different SF efficiencies 
(in $Gyr^{-1}$) assuming a formation redshift 5 (thin lines). It is 
also shown a
model with formation redshift $z_f$ = 8 and $\nu\,=\,0.1\,Gyr^{-1}$
(thick line). The other lines are the same as in Figure 12. 
The letters correspond to DLAs with both ratios observed.}
\end{figure}


In Figures 12 and 13 the predictions of the standard model for 
dSph galaxies are compared to [Si/Fe] corrected for dust depletion
and [N/Si] observed in DLAs as a function of metallicity and 
redshift, respectively. The corrected values of [Si/Fe] and [Fe/H]
are those from the case E00 of Vladilo (2002) and the
[N/Si] ratios were taken from the compilation of Centurion et al. 
(2003). The letters correspond to systems where both ratios are 
available. The appearance of these plots is however misleading and 
some remarks must be made. In the plot of [Si/Fe] and [N/Si] as 
a function of metallicity the dSph model seems to account for 
almost all the systems, but some of them, especially the 
ones at high metallicities ([Si/H] $>\,\sim$ -0.7),
achieve their  observed [N/Si] ratios 
only a long time after the onset of the wind. As
the wind is very efficient in removing the gas content of the 
galaxy, it is unlikely that the dSph galaxies could be related to 
these DLAs, since these objects contain a large amount of gas.
In fact, at the highest [Si/H] the models predict a HI column 
density below the ones characteristic of DLAs (log N(HI) 
$\leq\,20$ at [Si/H] $\geq$ -0.7 for the models 
with $\nu \leq 0.05\, Gyr ^{-1}$). All the other DLAs, however, could 
be explained as possible progenitors of dSph galaxies since there 
is a good agreement between the abundance ratios observed in DLAs
and the predictions of the dSph model.
A more complete comparison, though, could be made with the
systems which have both ratios measured. Unfortunately, there
are only 5 systems with simultaneous [Si/Fe] and [N/Si] (letters
A to E in Figures 12 and 13) and no accurate conclusions can
be reached. Only 2 out of 5 systems (B and C) seems to be 
reproduced by the models with the same SF efficiency and
at the same epoch. The other systems (A, D and E) require either 
a different SF efficiency or different epochs of burst occurrence
to reproduce each 
abundance ratio. 
In the diagram of abundance ratios as a function of redshift,
on the other hand, the lack of agreement between the 
predictions of the dSph models and the observed data does not mean
that a scenario where the 
DLAs become dSph galaxies should be discarded. For the sake 
of simplicity we assumed the same formation redshift for all 
systems, and this could be unrealistic. Each DLA could have been 
formed at a different cosmic epoch and have followed a
different evolutionary track. Therefore, this diagram shows that, if 
the dSph galaxies are the progenitors of the DLAs, one is seeing 
the earliest stages of evolution of these systems before or at 
the onset of the wind and that the dSph galaxies would be seen
as a DLAs only during a well defined period of time, no longer than 
4 Gyr. After that time, their HI column densities fall below the limit 
which characterizes the DLAs. A point which plays against 
the scenario
in which the DLAs would be the progenitors of the local dSph galaxies
is the differences in the abundance patterns of Zn observed in
DLAs and in dSph galaxies, respectively. 
In the dSph galaxies of the Local Group,
the observed Zn/Fe is always below solar (Shetrone, Cot\`e $\&$
Sargent 2001) whereas even in the less dusty DLAs this ratio is 
above solar. This fact might indicate that the DLAs are in fact
not related to dSph galaxies or that the evolution with time of
Fe and Zn are different or even that different amounts of Zn and Fe
are lost in the galactic wind. Unfortunately, the nucleosynthesis of 
Zn is not fully understood and no firm conclusions can be drawn.

\begin{figure}
\centering
\epsfig{file=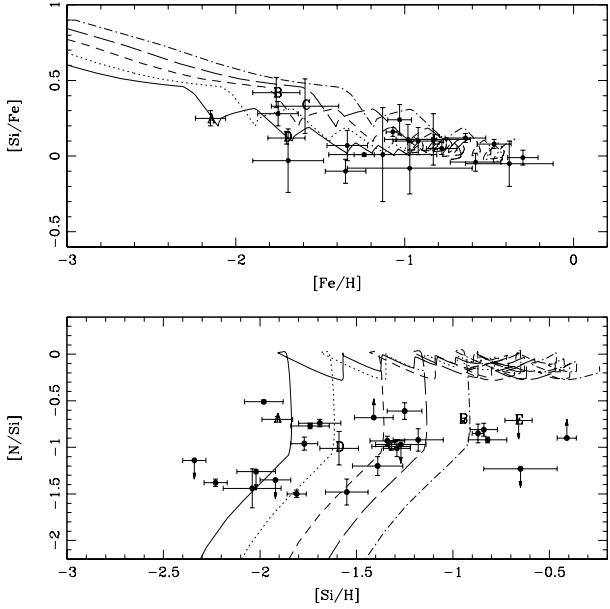,height=8cm,width=8cm}
\caption[]{[Si/Fe] vs [Fe/H] corrected for dust (above) and 
[N/Si] vs [Si/H] (below) observed in DLAs compared to the 
predictions of the standard model for BCGs with 7 bursts with 
different SF efficiencies (in $Gyr^{-1}$): 
0.3 (solid line), 0.5 (dotted), 0.9 (dashed), 1.5 (long-dashed),
 2.5 (dotted-dashed). The letters correspond to DLAs with both 
ratios observed.}
\end{figure}

\begin{figure}
\centering
\epsfig{file=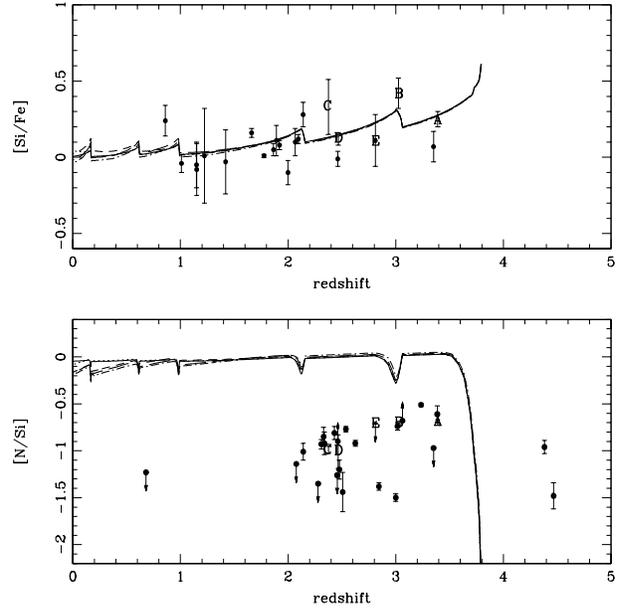,height=8cm,width=8cm}
\caption[]{Evolution with redshift of [Si/Fe] corrected for dust (above)
and [N/Si] (below) observed in DLAs compared to the predictions of the
standard model for BCGs with 7 bursts with different SF efficiencies 
(in $Gyr^{-1}$) assuming a formation redshift 5. The lines are 
the same as in Figure 14. The letters correspond to DLAs with both 
ratios observed.}
\end{figure}

The case of the BCGs is more promising. In
Figures 14 and 15, respectively, the predictions of the standard 
model with 7 bursts for BCGs compared to the observed [Si/Fe] 
(corrected for dust) and [N/Si] as a function of metallicity and 
redshift are shown, respectively. In all cases the models can account
for all the observed systems, if higher SF efficiencies are adopted.
While the data of BCGs require a SF efficiency in the range
$\nu\,=\,0.1-0.9\, Gyr^{-1}$, the abundance ratios in DLAs are
fully reproduced by the standard model for BCGs with $\nu$ in 
the range $\nu\,=\,0.3-2.5\, Gyr^{-1}$.  
In this case, the DLAs with both abundance ratios measured are
reproduced by the same models at the same time with the exception
of system E, which is not reproduced even by the dSph model.
In the BCGs scenario, contrary to the dSphs, only one formation
redshift ($z_f=5$) is necessary to reproduce the data at all 
epochs suggesting that the DLAs could have started forming stars at 
the same cosmic epoch and, be observed at different stages 
of evolution.

\begin{figure}
\centering
\epsfig{file=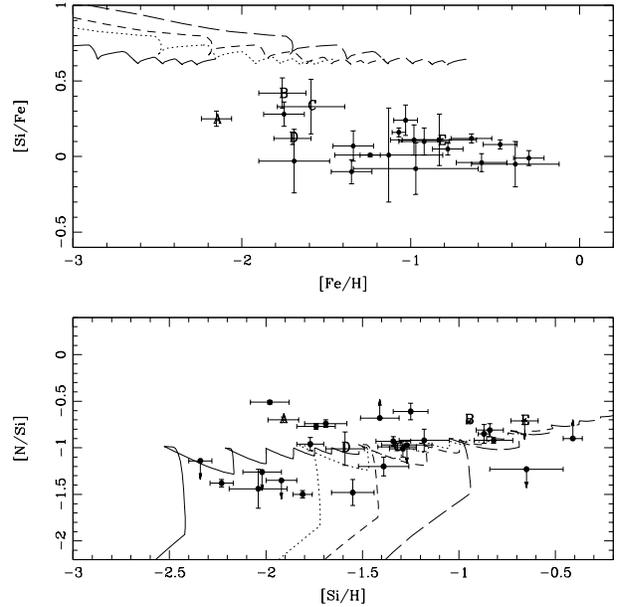,height=8cm,width=8cm}
\caption[]{[Si/Fe] vs [Fe/H] (above) corrected for dust and 
[N/Si] vs [Si/H] (below) observed in DLAs compared to the 
predictions of the standard
model for BCGs assuming a IMF with an exponent x=0.1. The lines 
represent different SF efficiencies (in $Gyr^{-1}$): 
0.01 (solid line), 0.05 (dotted), 0.1 (dashed), 0.3 (long-dashed).
The letters correspond to DLAs with both ratios observed.}
\end{figure}
\begin{figure}
\centering
\epsfig{file=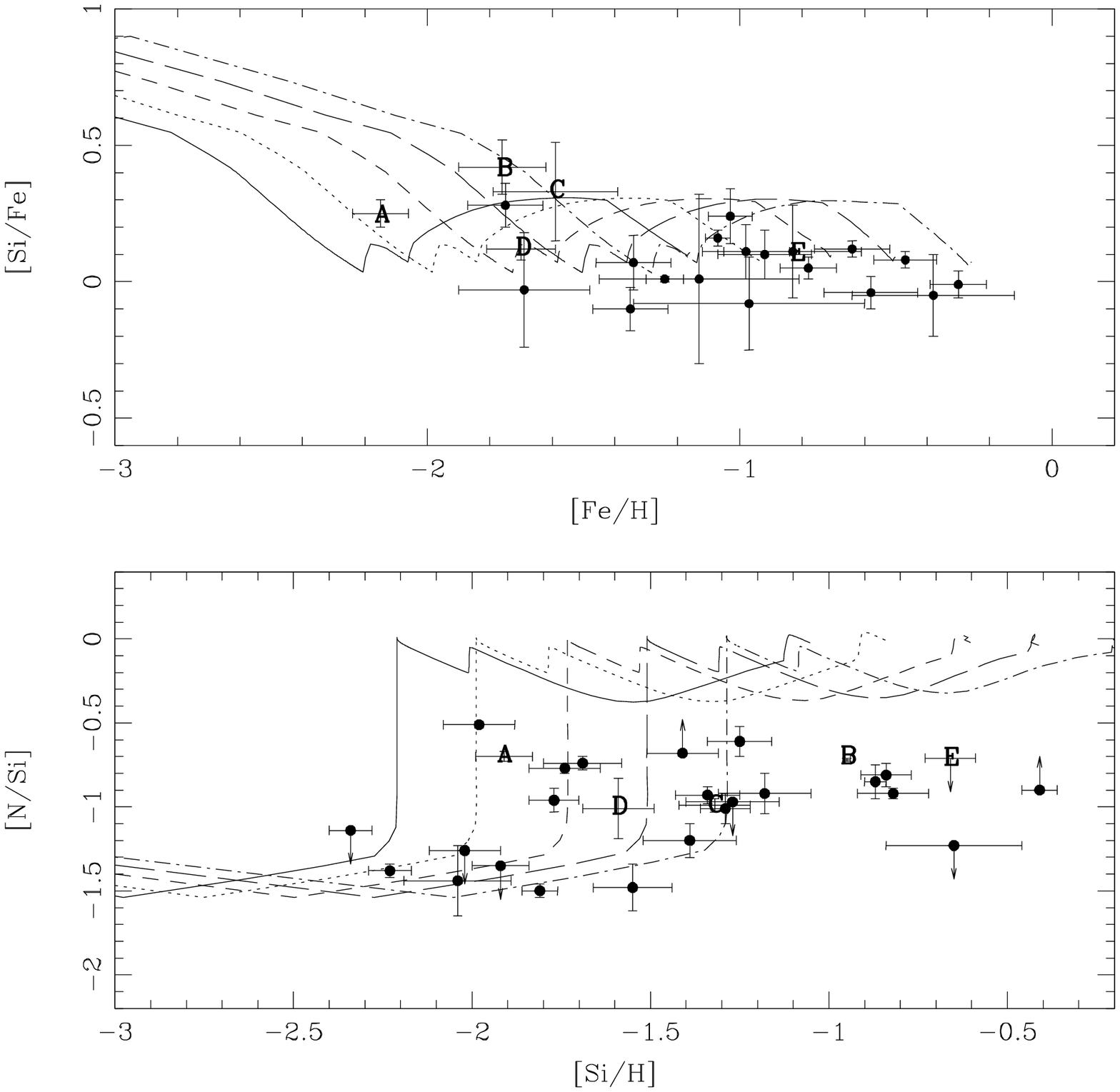,height=8cm,width=8cm}
\caption[]{[Si/Fe] vs [Fe/H] (above) corrected for dust and 
[N/Si] vs [Si/H] (below) observed in DLAs compared to the predictions of the standard
model for BCGs with 4 bursts assuming a primary production of N in 
massive stars. The lines represent different SF efficiencies
(in $Gyr^{-1}$): 0.3 (solid line), 0.5 (dotted), 0.9 (dashed), 
1.5 (long-dashed), 2.5 (dotted-dashed). The letters 
correspond to DLAs with both ratios observed.}
\end{figure}

Two remaining problems, though, are the high values for the 
predictions of [N/Si] and the claimed bimodal distribution
of DLAs (Prochaska et al. 2002; Centurion et al. 2003). We 
adopted two alternative scenarios in the BCGs model in order to 
try to explain these features. First we considered a top-heavy
IMF with an exponent x=0.1, as suggested by Prochaska et al. (2002)
and Henry $\&$ Prochaska (2003). Indeed, if a top-heavy IMF is 
assumed the models predict very low values for [N/Si], but, at the
same time, very high [Si/Fe] ratios, well above all systems 
observed (Figure 16), making clear that a top-heavy IMF cannot 
explain the low values found for some DLAs. Another possibility 
would be the model for BCGs with 4 bursts, since it predicts low 
values of N/O during the third burst. We used this model 
and included primary production of N in massive 
stars in order to try to explain 
the plateau at low values of [N/$\alpha$] suggested by Prochaska et 
al. (2002) and Centurion et al. (2003). If a very low amount 
of primary N, roughly in agreement with the yields of the 
model with Z=0.004 of Meynet $\&$ Maeder (2002), is assumed to 
be produced in massive stars, then the models
reproduce the N/O distribution of BCGs (Figure 10) and 
predict a plateau at the same values observed in low-N DLAs (Figure 
17). This plateau, in the models, is a result of a very short burst 
(20 Myr) of SF at the very early stages of the evolution
combined with the assumed low level of primary production of N
in massive stars. During the burst both $\alpha$ elements and N are
produced in massive stars and injected into the ISM giving rise to a 
constant N/$\alpha$ ratio, which has a low value due to the short 
burst duration and to the low amount of primary N produced. 
It should be mentioned, however, that the suggested bimodal
distribution, characterized by this plateau at low values of 
[N/$\alpha$], could not be real and should be confirmed by more data,
and that the production of primary N in massive stars, altough 
predicted by stellar models with rotation (Meynet $\&$ Maeder 2002), 
is not necessarily required to explain the N/O ratio observed in 
different environments (Chiappini, Romano $\&$ Matteucci 2003).

\section{Summary}

We analysed the star formation and chemical evolution both in
dSph galaxies of the Local Group and in 
BCGs comparing several observed abundance 
ratios to the predictions of detailed chemical evolution models.
By taking into account the role played by supernovae of 
different types (II, Ia) and adopting up to date nucleosynthesis
prescriptions we followed the evolution of several chemical elements 
(H, D, He, C, N, O, Mg, Si, S, Ca, and Fe). Each galaxy model was 
specified by the prescriptions of the SF and wind efficiency in 
the sense that one long episode of star formation and a high wind 
efficiency characterize the dSph galaxies, whereas the BCGs are
represented by SF occurring in several short bursts separated by 
long quiescent periods and a low wind efficiency. The predictions 
of the models were compared with the [$\alpha$/Fe] ratios in dSph 
galaxies and with C/O, N/O, Si/O and with [O/Fe] in BCGs. A possible 
connection between these two types of galaxies in an unified 
evolutionary scenario as well as the hypothesis that the dSph 
galaxies or BCGs could be related to DLAs were also analysed.

The main conclusions can be summarized as follows: 

\begin{itemize}

\item
by adopting a standard model for the sample of 8 dSph galaxies of 
the Local Group, the observed [$\alpha$/Fe] ratios are reproduced 
with a SF occurring in 1 long burst lasting 8 Gyr, with an efficiency of
SF in the range $\nu=0.01-1.0 \,Gyr^{-1}$; 

\item
each galaxy with enough data is also reproduced by a model following
the main prescriptions of the standard model, but with the number,
epoch and duration of the bursts as suggested by the SF histories inferred
by CMDs. The differences in the abundance patterns of each galaxy are 
mainly due to different efficiencies in the star formation rate rather than 
to details in the histories of SF;

\item
a very efficient galactic wind ($w_i \sim 5 -15$, 
depending on the SF efficiency) is required 
to reproduce the gas content, total mass and abundance ratios
observed in local dSph galaxies. The low gas content of the 
dSph galaxies is the result of the consumption of gas by the SF 
coupled with gas removal by galactic winds;

\item
the N/O, C/O, Si/O and [O/Fe] ratios observed in BCGs can be 
explained by a model with 2 to 7 short bursts of SF with
efficiencies in the range $\nu=0.1 - 0.9\, Gyr^{-1}$; 

\item
the low values of N/O at low O/H observed in BCGs could be 
the natural result of a model with 4 bursts of SF where the 
first one occurs at early galactic ages and the others at more 
recent times. In such a scenario, there is no need of invoking 
primary production of N in massive stars, even though the adoption
of low amounts of primary N in these stars does not change much
the results;

\item
a Salpeter (1955) IMF seems to reproduce at best the properties of
both dSph galaxies and BCGs;

\item
a connection between dSph galaxies and BCGs in an unified 
evolutionary scenario is not likely, owing to the large 
differences in the chemical evolution history of these galaxies;

\item
the dSph and BCG models suggest two different possible pictures 
for the DLAs: while the BCG models can reproduce all observed DLAs
at all epochs and suggest that these systems could have been formed
at the same cosmic epoch (around $z_f$ = 5) but observed at 
different stages of their evolution, only the DLAs with low 
metallicities ([Si/H] $\leq$ -0.7) could be related to dSph 
galaxies and, in this case, they should be very young systems 
which have been formed within a wide range of redshifts;

\item
the suggested plateau at low N/$\alpha$ observed in DLAs can
possibly be explained by the model for BCGs with 4 bursts of SF
if a low level of primary N produced in massive stars is adopted.
In this case, the plateau should be the result of a very short 
(20 Myr) initial starburst when both $\alpha$-elements and N would
be produced by massive stars. As the burst is short and the amount
of primary N is low, the predicted value of the N/$\alpha$ ratio 
would be also low. The predicted $\alpha$/Fe ratios, however, would
be very high ([$\alpha$/Fe] $\geq$ 0.5), above all the observed 
values and at metallicities well below the ones seen in DLAs
([Fe/H] $\leq$ -2.3). Consequently, if this scenario
is the explanation for the suggested plateau, then one should 
observe such low Fe abundances and high $\alpha$/Fe ratios in
the DLAs which have low N/$\alpha$ ($\sim$ -1.6 dex). Another 
uncertainty concerning the 
plateau predicted by the models is the adopted amount of primary N
produced in massive stars. Our value is roughly in agreement with 
the model with Z=0.004 of Meynet $\&$ Maeder (2002),
but only the BCG model with the highest SF efficiency reaches
this metallicity. For all the other models, the amount of primary N 
produced by massive stars would be lower (as in the model with
Z=$10^{-5}$ of Meynet $\&$ Maeder 2002) and consequently the 
predicted N/$\alpha$ would lie below the observed plateau. 
Therefore, in this case, the primary N produced by massive stars,
as suggested by the yields of Meynet $\&$ Maeder (2002), could not
be the explanation for the observed plateau, if this plateau is real.

\end{itemize}

\section*{Acknowledgments}
We thank C. Chiappini, F. Calura, S. Recchi for many useful discussions.
G.A.L. also thanks the hospitality of the Dipartimento di 
Astronomia-Universit\'a di Trieste and of the Astronomical Observatory
of Trieste which provided all the support
during his stay.
G.A.L. acknowledges financial support from the Brazilian agency 
FAPESP (proj. 00/10972-0).

\bsp

\end{document}